\documentclass[journal]{IEEEtran}
\usepackage{hyperref}
\usepackage[utf8]{inputenc}
\usepackage[utf8]{inputenc}
\usepackage{graphicx}
\usepackage{amsmath}
\usepackage{amsthm}
\usepackage{amssymb,stmaryrd,amsmath,amsfonts,rotating}
\usepackage{amsmath, amsthm, amssymb}
\usepackage{epstopdf}
\usepackage[mathscr]{eucal}  % associates Euler fonts with \mathscr (default associates with \mathcal
\usepackage{float} 
\usepackage{multicol}
\usepackage{caption}
\usepackage{subcaption}
\usepackage{enumerate}
\usepackage{authblk}
\usepackage{multirow}
\usepackage{tikz}
\usetikzlibrary{decorations.pathreplacing}
\usepackage{multirow}
\usepackage{cite}

\usetikzlibrary{positioning}

\usepackage{tikz}
    \usetikzlibrary{shapes.geometric}   
\usepackage{rotating}

\usepackage{stmaryrd}
\usepackage{mathtools}
\usepackage{stfloats}
\usepackage{url}
\usepackage{amssymb}
\usepackage{mathabx}
\usepackage{algorithmicx}
\usepackage{physics}

\newcommand{\defn}{\stackrel{\Delta}{=}}

\newtheorem{proposition}{Proposition}
\newtheorem{corollary}{Corollary}
\newtheorem{theorem}{Theorem}

\renewcommand{\theenumi}{\roman{enumi}}
\renewcommand{\vec}{\vb*}

\begin{document}


\title{Statistical Classification via  Robust Hypothesis Testing: Non-Asymptotic and Simple Bounds }
\author{H\" useyin~Af\c ser \thanks{ \hspace{- 11.2pt} Submission date is 27/08/2021. \newline \hspace{18pt} H\" useyin~Af\c ser is with the Adana Alparslan T\"{u}rke\c{s} Science and Technology University, Department  of Electrical Electronics Engineering, 01250, Adana, Turkey. (e-mail:afser@atu.edu.tr)}}

\maketitle

\begin{abstract}

We consider Bayesian multiple statistical classification problem in the case where the unknown source distributions are estimated from the labeled training sequences, then the estimates are used  as nominal distributions in a robust hypothesis test. Specifically, we employ the DGL test due to Devroye et al. and provide non-asymptotic, exponential upper bounds on the error probability of classification. The proposed upper bounds are simple to evaluate and reveal the effects of the length of the training sequences, the alphabet size and the numbers of  hypothesis on the error exponent. The proposed method can also be used for large alphabet sources when the alphabet grows sub-quadratically in the length of the test sequence.
The simulations indicate that the performance of the proposed method gets close to that of optimal hypothesis testing as the length of the training sequences increases.

\end{abstract}


\section{Introduction}

In classical multiple hypothesis testing the aim is to choose between $M$ sources, with known distributions, that is responsible
for the generation of an observed test sequence \cite{Kay}. Statistical classification addresses the same problem with the only distinction that
the source distributions are not known exactly, but one has at his disposal labeled training sequences that are generated by each source \cite{Mitchie}.

In  Bayesian hypothesis testing one assumes positive
prior probabilities for the hypothesis and the optimal test is the maximum a posteriori (MAP) decision rule.  The corresponding error exponent is the minimum pairwise Chernoff distance between distinct source distributions \cite{Leang}. In  Neyman-Pearson setting, one seeks a trade-off by maximizing the error exponent of a single hypothesis while ensuring that the remaining error probabilities do not exceed a prescribed threshold. Tests based on Chernoff-Stein lemma are optimal and the corresponding $M(M-1)$ error exponent trade-off region is characterized by Tuncel \cite{Tuncel}.

Intuitively, as the length of the training sequences gets very large, their empirical distributions converge to true ones and the problem becomes identical to hypothesis testing. In this asymptotic region, Cover investigated the performance of the nearest neighbour decision rule and showed that the classification error of a single observation is bounded by twice that of Bayesian hypothesis test \cite{Cover}. In \cite{Zhou}, authors investigated the binary classification problem in Neyman-Pearson setting and obtained second order (dispersion type) error upper bounds for the test originally proposed by Gutman in \cite{Gutman}. They showed that that this test is second order asymptotically optimal for any scaling on the length of the training sequences. In \cite{Haghifam} authors proposed a sequential test and showed that it performs better than Gutman's test in terms of  Bayesian error exponent. Kelly et al. \cite{Kelly} considered the binary classification problem for large alphabet sources and formalized the maximum growth rate of the alphabet for asymptotically consistent classification. Huang et al. \cite{Huang} investigated the same problem by considering different lengths for the test and the training sequences.

In practical applications, obtaining labeled training sequences is time consuming and cumbersome. Thus, it is crucial to characterize the performance of classification such that the role of training sequences becomes evident at finite and practical lengths. Motivated by this fact and the lack of studies on Bayesian multiple classification in the non-asymptotic region, we propose to use robust hypothesis testing \cite{HuberB,HuberA, Levy} where the nominal (estimate) distributions are used instead of true ones and the test is robust to small deviations between them. Specifically, we  employ the DGL test due to Devroye et al. \cite{Devroye, Biglieri}  since this test can be used for multiple classification and it has a non-asymptotic, exponential upper bound on its error probability.

The contributions of this paper are as follows: We extend the non-asymptotic, exponential upper bound of the DGL test for the considered problem in a way that the effect of the length of the training sequences on the error exponent becomes evident. We show, via simulations, that the performance of the proposed method gets close to that of optimal hypothesis testing at practical lengths. The proposed upper bounds are simple to evaluate and also provide insight on the effect of the number of hypothesis and the source alphabet size on the error exponent. In this regards, we investigate the large alphabet case and show that the proposed method can be used even when the alphabet size grows faster than the length of the test sequence. Our work contributes to the existing literature \cite{Cover, Zhou,Gutman,Haghifam,Kelly, Huang} by considering the non-asymptotic region, and complements the large alphabet case \cite{Kelly, Huang} with findings on multiple classification.  
 
The outline of the paper is a follows: In Section \eqref{sect:sect_2}, we present the preliminary material,  explain the classification problem and review the DGL test in the discrete setting. In Section \eqref{sect:sect_3}, we explain the proposed classification method and derive upper bounds on its error probability. Then, we refine the bounds for large alphabet sources and provide simulations.  Finally, Section \eqref{sect:sect_5} concludes the paper.

\section{Preliminaries}\label{sect:sect_2}

\subsection{Statistical Distances}
 Let $P$ and $Q$ be two distributions defined over a common, discrete alphabet $ \cal X$. The total variation (distance) between $P$ and $Q$ is defined as
\begin{align}
V(P,Q) &= \max_{ \hspace{10 pt} F, F \subset \cal{X}} |P(F) -Q(F)|, \label{eq:V_def1} \\
	   & = \frac{1}{2}\sum_{a \in \cal X} |P(a)-Q(a)|. \label{eq:V_def2}	
\end{align}
Chernoff distance  between $P$ and $Q$ is 
\begin{align}
C(P,Q) = -\underset{\lambda \in [0,1]}{ \text{min}} \log_2 \left( \sum_{a \in \mathcal{X}} P(a)^\lambda Q(a)^{1-\lambda} \right).
\end{align}
Following is an inequality between Chernoff distance and total variation \cite[Corollary 4]{Sason}.
\begin{align}
C(P,Q) \geq -\frac{1}{2} \log_2 (1 -V(P,Q)^2). \label{eq:Sason_ineq}
\end{align}

\subsection{Multiple Classification Problem}
We need to classify a sequence $\vec{x}^n=[x_{1},x_{2},....,x_{n}]$, $\vec{x}^n \in {\cal X}^n$, where each observation in $\vec{x}^n$ is independent and identically distributed (i.i.d) from one of $M$ possible sources. In this paper we assume that the source alphabet ${\cal X}$ is discrete and countably finite. The true distributions of the sources, $P_1,P_2,...,P_M$, are not known; however  there exists training sequences $\vec{t}_i^N=[ t_{i1},t_{i2},....,t_{iN} ]$, $ i=1,2,....,M$, and it is known that $\vec{t}_i^N$ has emerged from source $i$.  If we define ${\cal H}_i$ to be the hypothesis that $\vec{x}^n$ is generated by source $i$, then the aim is to come up with a decision rule $|{\cal X}|^{n+MN} \rightarrow  \{ \Omega_1,\Omega_2,...,\Omega_M \}$ such that ${\cal H}_i$ is selected if  $[\vec{x}^n,\vec{t}_1^N,\vec{t}_2^N,...,\vec{t}_M^N]\in \Omega_i$.

\subsection{DGL Test}

The DGL test \cite{Devroye} is a robust multiple hypothesis testing procedure for i.i.d. sequences. It 
can be used when the true distributions of the hypothesis are not known, but
there exist nominal distributions, $T_1,T_2,...,T_M$, that are close to true ones in total variation
 \footnote{The DGL test can be used when the underlying alphabet is continuous as well. In this letter, we review the DGL test by assuming that the source alphabet is discrete and countably finite.}. Assume that we want to test whether a sequence is generated according to ${\cal H}_i$. In this scenario, the test is robust provided that there exists 
a positive $\Delta$ such that  
\begin{align}
V(P_i,T_i) \leq (\min_{j, \hspace{1 pt}j \neq i} V(T_i,T_j) -\Delta)/2. \label{eq:DGL_robustnes}
\end{align}
Upon observing the test sequence, $\vec{x}^n$, one calculates the statistics $\mu_n( A) = \frac{1}{n} \sum_{i=1}^{n} I_{x_i \in  A}$,
where $I$ is an indicator function and $A$ is a borel set. Let $\cal{A}$ denote the collection of 
$M(M-1)/2$ sets that are of the form
\begin{align}
A_{i,j} = \{ a : T_i(a) \geq T_j(a) \},  \quad 1 \leq i < j \leq M.
\end{align}
The test accepts $H_i$ if
\begin{align}
\hspace{-8 pt}\max_{A \in \cal{A}} \sum_A |T_i-\mu_n( A) | = \min_{j}  \max_{A \in \cal {A} }  \sum_A |T_j-\mu_n( A) |. \label{eq:DGL_test_rule}
\end{align}
 Let $\Pr [e_{\textnormal{DGL}}]$ denote the resultant probability of error, averaged over $\vec{x}^n, \vec{x}^n \in {\cal X}^n$. 
$\Pr [e_{\textnormal{DGL}}]$ obeys a non-asymptotic, exponential upper bound of the form \cite{Devroye}
\begin{align}
\Pr [e_{\textnormal{DGL}}] \leq 2M e^{-n( \frac{\Delta^2}{2}-\frac{2 \ln (M-1)}{n})}. \label{eq:dgl_bound}
\end{align}
This upper bound is uniform in the sense that it does not depend on the particular $T_i$ and holds for testing $\cup_{i=1}^M {\cal H}_i$ provided that $\eqref{eq:DGL_robustnes}$ is satisfied.
In a practical implementation, the sets in $\cal{A}$ can be calculated prior to the test. Then, \eqref{eq:DGL_test_rule} can be performed with complexity 
$O(M^2n+M^2 \log M)$ \cite{Devroye}.
\vspace{-10 pt}
\section{ DGL Test-Based Classification}\label{sect:sect_3} 
\subsection{Implementation and Performance}\label{sect:sect_3A}
%\begin{figure*}
%\begin{center}
%\input{fig_1}
%%\includegraphics[scale=0.4]{fig_1.png}
%\end{center}
%\caption{The geometry of the proposed classification method conditioned on the event $\Phi$.}
%\label{fig:fig_1}
%\end{figure*}

Given the training sequences, $\vec{t}^N_i$, $i=1,2,...,M$, one can use the DGL test by choosing the nominal distribution, $T_i$, as the empirical distribution of the training sequence  $\vec{t}^N_i$. Then, the classification task can be carried out with the test in $\eqref{eq:DGL_test_rule}$ and has the same complexity as the DGL test.

In order to investigate the resultant classification method, let us define 
\begin{align}
\phi_i  &\defn \{ \vec{t}^N_i : V(T_i,P_i) \leq  \frac{\min_{j, \hspace{1 pt} j \neq i} V(T_i,T_j) -\Delta}{2} \}, \\
\Phi &\defn \cap_{i=1}^M \phi_i.
\end{align}
Conditioned on the event $\Phi$, the DGL test is robust for testing $\cup_{i=1}^M {\cal H}_i$. If we let $\Pr [e_{\textnormal{CL}}]$ denote the error probability of classification, $\Pr [e_{\textnormal{CL}} | \Phi] = \Pr [e_{\textnormal{DGL}}]$ holds. Furthermore
\begin{align}
\Pr [e_{\textnormal{CL}}] &= \Pr [e_{\textnormal{CL}}|\Phi]\Pr [\Phi] + \Pr [e_{\textnormal{CL}}|\Phi^c]\Pr [\Phi^c], \nonumber\\
                 &\leq \Pr [e_{\textnormal{DGL}}] +  \Pr [\Phi^c]. \label{eq:CL_ineq_1}
\end{align}
The term $\Pr [\Phi^c]$ can be regarded as the estimation error that occurs when 
that the nominal (estimate) distributions are not close to true distributions as required for the robustness of the DGL test. We know that $ \Pr [e_{\textnormal{DGL}}]$ is exponentially decaying in $n$, and the law of large numbers \cite[Thm. 11.2.1]{Cover} implies
that  $\Pr [\Phi^c]$ is exponentially decaying in $N$. $\Pr [e_{\textnormal{CL}}]$ is dominated by the minimum of the exponents of $ \Pr [e_{\textnormal{DGL}}]$ and $\Pr [\Phi^c]$, thus we seek the point where the two exponent are equal. We let $N$ to scale with $n$ and define  
\begin{align}
\alpha \defn N/n.
\end{align}
The following proposition is proved in the Appendix.
\begin{proposition}\label{thm:prop_1}
\begin{align}
\Pr[ \Phi^c]   \leq 2M e^{-n \left(\frac{  \alpha  \min_{i \neq j} (V(T_i,T_j) -\Delta)^2}{2} - \frac{| {\cal X} | \ln 2}{n} \right) }.  \label{eq:phi_bound}
\end{align}
\end{proposition}
Using \eqref{eq:dgl_bound} and \eqref{eq:phi_bound} in \eqref{eq:CL_ineq_1} results in
\begin{align}
\Pr [e_{\textnormal{CL}}] &\leq 2M e^{-n( \frac{\Delta^2}{2}-\frac{2 \ln (M-1)}{n})}+ \nonumber \\ 
&  \quad \quad \quad 2M e^{-n \left( \frac{ \alpha \min_{i \neq j} (V(T_i,T_j) -\Delta)^2}{2} - \frac{| {\cal X} | \ln 2}{n} \right) }. \label{eq:phi_upper_bound}
\end{align}
The above upper bound is valid  $\forall \Delta$, $ \Delta > 0$. The leading coefficients of the exponents are equal when
\begin{align}
\Delta = \sqrt{ \alpha } \min_{i \neq j} V(T_i,T_j)/({1+\sqrt{\alpha}}. \label{eq:delta})
\end{align}  

Using  \eqref{eq:delta} in \eqref{eq:phi_upper_bound} results in a non-asymptotic, exponential bound on $\Pr [e_{\textnormal{CL}}]$. This is presented in the following Theorem.

\begin{theorem}\label{thm:thm_1}
\begin{align}
\hspace{-11 pt} \Pr [e_{\textnormal{CL}}]   \leq 2Me^{-n ( \frac{ \alpha \min_{i \neq j}V(T_i,T_j)^2 }{2(1 + \sqrt{\alpha})^2} - \max \{ \frac{2 \ln (M-1)}{n}, \frac{| {\cal X} | \ln 2}{n} \} )}\hspace{ -4 pt}.
\hspace{-8 pt} 
\end{align}
\end{theorem}

Theorem \ref{thm:thm_1} is useful for a data driven analysis in the sense that one can obtain an upper bound on the classification error by using the empirical distributions of the training sequences. Next, we try to relate the upper bound in Theorem \ref{thm:thm_1}  to the unknown true distributions of the the sources which will help us to make connections with the existing results in the literature.  We obtain this relationship by deriving an inequality between $\min_{i \neq j}V(T_i,T_j)$ and $\min_{i \neq j}V(P_i,P_j)$.  This is presented in the following proposition whose
proof is provided in the Appendix.
\begin{proposition}\label{thm:prop_2}
For some arbitrary $\epsilon(\alpha)$, $\epsilon(\alpha)>0$, conditioned on $V(T_i,P_i) \leq \min_{i \neq j} V(T_i,T_j) \epsilon(\alpha)$, $i=1,2,...,M$,
\begin{align}
\min_{i \neq j} V(T_i,T_j) \geq  \frac{ \min_{i \neq j} V(P_i,P_j)}{1 + 2\epsilon(\alpha)}.
\end{align}
\end{proposition}

Notice that conditioned on $\Phi$ and when $\Delta$ is of the form \eqref{eq:delta},  $V(T_i,P_i) \leq \min_{i \neq j} V(T_i,T_j) \epsilon(\alpha)$ holds with $\epsilon(\alpha)= \frac{1}{2(1+\sqrt{\alpha})}$. In turn, Proposition \eqref{thm:prop_2} implies $\min_{i \neq j} V(T_i,T_j) \geq \min_{i \neq j} V(P_i,P_j) (\frac{1+\sqrt{\alpha}}{2+\sqrt{\alpha}}).$ Using this fact in Theorem \eqref{thm:thm_1} results in the following corollary.

\begin{corollary}\label{thm:cor_1}
\begin{align}
\hspace{-11 pt} \Pr [e_{\textnormal{CL}}]  \leq 2Me^{-n ( \frac{ \alpha \min_{i \neq j}V(P_i,P_j)^2 }{2(2 + \sqrt{\alpha})^2} - \max \{ \frac{2 \ln (M-1)}{n}, \frac{| {\cal X} | \ln 2}{n}\} )} \hspace{-4 pt}.\hspace{-8 pt} \label{eq:upper_bound_2}
\end{align}
\end{corollary}

When $\max \{ \ln (M-1), | {\cal X} | \}$ scales slower than $n$, Theorem \ref{thm:thm_1} and Corollary \ref{thm:cor_1} indicate that the error exponent scales with $\min_{i \neq j}V(T_i,T_j)^2$ and $\min_{i \neq j}V(P_i,P_j)^2$, respectively. Thus, for any $\alpha >0$, the proposed method offers consistent classification provided that the unknown sources are separated in variational distance. We have
\begin{align}
\min_{i \neq j}V(P_i,P_j)^2/2 &\leq -\ln(1-\min_{i \neq j}V(P_i,P_j)^2)/2,  \label{eq:dgl_exponent} \\
		       &\leq \min_{i \neq j}C(P_i,P_j)	\ln 2,	 \label{eq:bayes_exponent}
\end{align}
where the first inequality results from $ \ln(z) \leq z-1$, $z \geq 0$, and the second one is due to \eqref{eq:Sason_ineq}. The
right hand side of \eqref{eq:bayes_exponent} is the achievable error exponent of Bayesian multiple hypothesis testing when $n$ is sufficiently large  \cite{Westover}.
Therefore, even when $\alpha$ is large, the exponent of the proposed upper bound in \eqref{eq:upper_bound_2} is less than the optimal exponent of hypothesis testing. This is expected
because the left hand side of \eqref{eq:dgl_exponent} is the error exponent of the original DGL test when nominal distributions are the same as true ones.
However, the proposed, non-asymptotic bound can be useful for lower bounding the achievable error exponent of classification for finite $n$ and $\alpha$.
 As an example, we have investigated the classification problem with $|{ \cal X}|=3$, $M=5$,  and the true distributions of the hypothesis are 
\begin{align*}
&P_1= [0.1,0.8,0.1],
P_2= [0.3,0.2,0.5], 
P_3= [0.6,0.1,0.3],  \\
& \quad \quad \quad \quad P_4= [0.4,0.4,0.2], 
P_5= [0.3,0.6,0.1].
\end{align*}
The simulation results are provided in Figure \ref{fig:fig_2} where the performance of the MAP decision rule is also included for comparison. We observe that the simulated $\Pr[e_{CL}]$ curves have negative slopes, i.e. positive error exponents, for $\alpha \in \{ 0.1, 1, 10,100 \}$. The proposed upper bound in \eqref{eq:upper_bound_2} is not tight when $\alpha$ is small; however, its slope gets closer to that of simulated error curves as $\alpha$ increases,. For the considered example, the performance of classification with $\alpha \in \{10,100 \}$ almost matches the performance of the MAP test. 

\begin{figure}
\begin{center}
\includegraphics[scale=0.22]{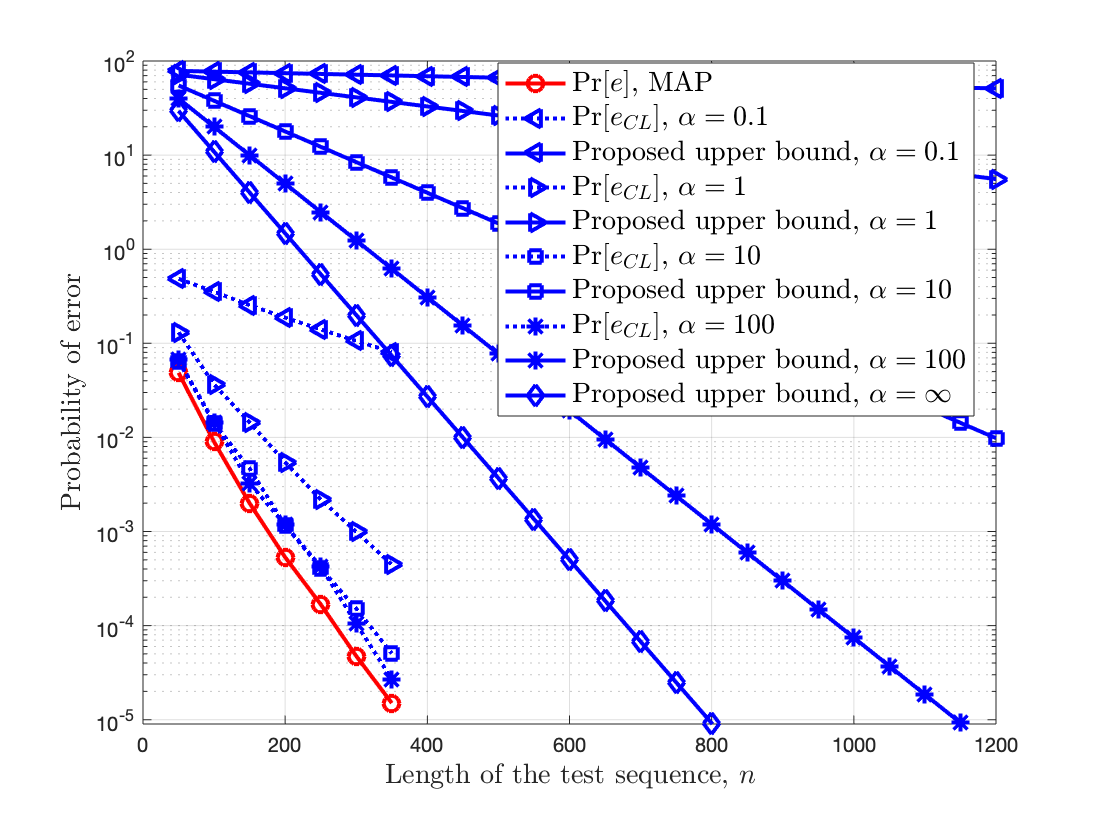}
\end{center}
\caption{Simulated error probabilities and the proposed upper bound in Corollary \eqref{thm:cor_1}.}
\label{fig:fig_2}
\end{figure}
\vspace{-10 pt}
\subsection{Large Alphabet Case}
The upper bound in Corollary \ref{thm:cor_1} implies that, for a positive error exponent, $|\cal X|$ can grow linearly in $n$ provided that
\begin{align}
|{\cal X}| < n  \left( \frac{  \alpha \min_{i \neq j}V(P_i,P_j)^2 }{2 \ln 2(2 + \sqrt{\alpha})^2} \right).  \label{eq:lin_scale}
\end{align}
However, the linear growth rate in \eqref{eq:lin_scale}  can not be improved by letting $\alpha \rightarrow \infty$.  This results from the fact that the upper bound on $\Phi^c$ in 
Proposition \ref{thm:prop_1} is not tight when $|{\cal X}|$ is comparable to $n$. Following proposition, whose proof
is provided in the Appendix, provides a remedy for this situation.
\begin{proposition}
\begin{align}
\Pr[ \Phi^c]   \leq 2M  e^{- n \left(  2\alpha (\frac{\min_{i \neq j} V(T_i,T_j) -\Delta}{ | {\cal X} |})^2 -\frac{\ln |\cal X|}{n}\right) }. \label{eq:phi_bound_2}
\end{align}
\end{proposition}
By following the same approach in Section \ref{sect:sect_3A} and using \eqref{eq:phi_bound_2} in \eqref{eq:CL_ineq_1} we obtain
\begin{align}
\Pr [e_{\textnormal{CL}}] &\leq 2M e^{-n( \frac{\Delta^2}{2}-\frac{2 \ln (M-1)}{n})}+ \nonumber \\ 
&  \quad \quad \quad 2M  e^{- n (  2\alpha (\frac{\min_{i \neq j} V(T_i,T_j) -\Delta}{ | {\cal X} |})^2 -\frac{\ln |\cal X|}{n} ) }. \label{eq:phi_upper_bound_2}
\end{align}
The two leading exponential terms above are equal to each other when 

\begin{figure}
\begin{center}
\includegraphics[scale=0.22]{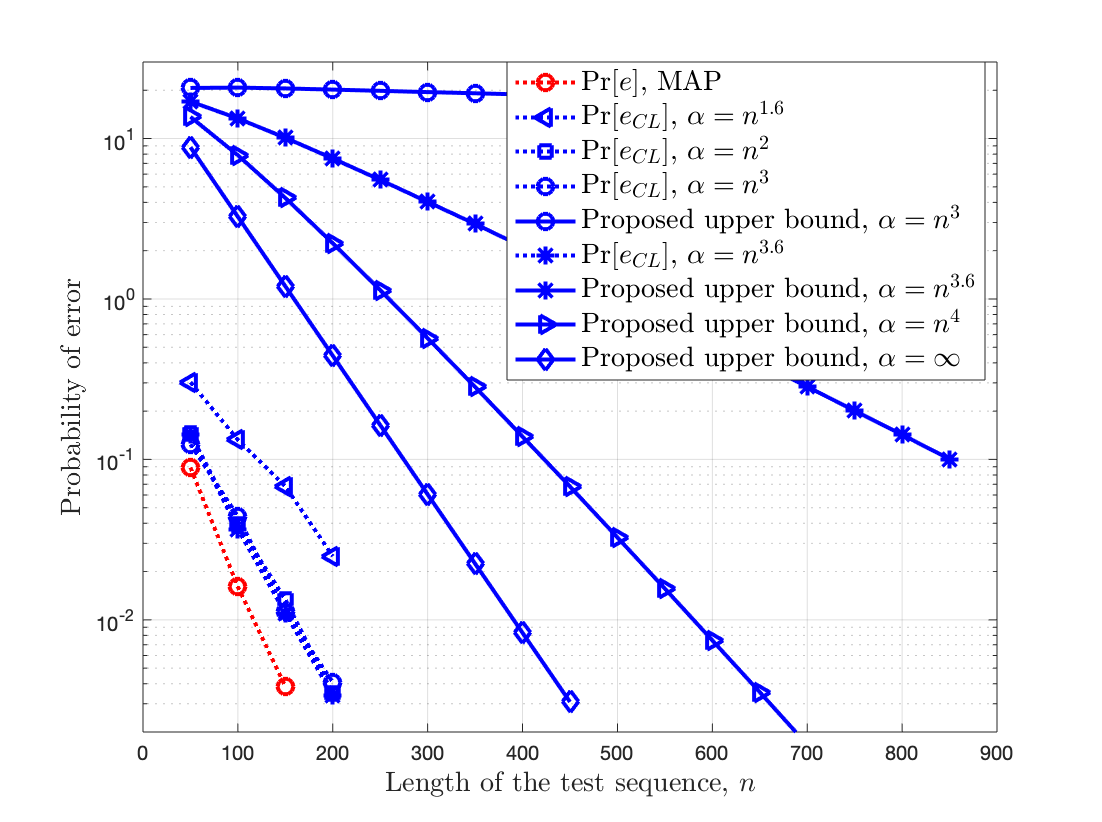}
\end{center}
\caption{Simulated error probabilities for the large alphabet case and the proposed upper bound in Corollary \eqref{thm:cor_2}.}
\label{fig:fig_2}
\end{figure}

\begin{align}
\Delta =   2\sqrt{\alpha} \min_{i \neq j}V(T_i,T_j) /({|\cal X|} +2\sqrt{\alpha} ).   \label{eq:delta_2}
\end{align}
which implies the following upper bound.
\begin{theorem}\label{thm:thm_2}
\begin{align}
\hspace{- 10 pt}\Pr [e_{\textnormal{CL}}]  \leq 2M e^{-n ( \frac{ 2\alpha \min_{i \neq j}V(T_i,T_j)^2}{{(| \cal X|}+2\sqrt{\alpha})^2}-  \max \{ \frac{2 \ln (M-1)}{n}, \frac{\ln |{\cal X}|}{n} \} ) }
\hspace{-4 pt}. \hspace{-8 pt}\label{eq:upper_bound_3}
\end{align}
\end{theorem}
Conditioned on $\Phi$ and when $\Delta$ is of the form \eqref{eq:delta_2} we have 
 $V(T_i,P_i) \leq \min_{i \neq j} V(T_i,T_j) \frac{|\cal X|}{|{\cal X}|+2 \sqrt{\alpha}}$. Then, letting $\epsilon(\alpha)=\frac{{|\cal X|}}{|{\cal X}|+2 \sqrt{\alpha}}$ in Proposition \ref{thm:prop_2}
one obtains $\min_{i \neq j} V(T_i,T_j) \geq \min_{i \neq j} V(P_i,P_j) (\frac{|{\cal X}|+2 \sqrt{\alpha}}{3|{\cal X}|+2 \sqrt{\alpha}})$. By using this fact in Theorem \ref{thm:thm_2} we obtain the following corollary.

\begin{corollary}\label{thm:cor_2}
\begin{align}
\hspace{- 10 pt}\Pr [e_{\textnormal{CL}}]  \leq 2M e^{-n ( \frac{ 2\alpha \min_{i \neq j}V(P_i,P_j)^2}{{(3| \cal X|}+2\sqrt{\alpha})^2}-  \max \{ \frac{2 \ln (M-1)}{n}, \frac{\ln |{\cal X}|}{n} \} ) }
\hspace{-4 pt}. \hspace{-8 pt}
\end{align}
\end{corollary}
When $M$ and $|{\cal X}|$ is not exponential in $n$, Corollary \ref{thm:cor_2} implies a non-vanishing error exponent provided that $| \cal X|$ scales slower than $\sqrt{\alpha}$. Therefore, the proposed DGL test-based method can be used even when 
$|\cal X|$ grows faster than $n$. This provides a performance advantage over chi-square and generalized likelihood ratio tests because these test can only provide sub-linear growth for $|{\cal X}|$ \cite{Kelly}. We would like to note that the growth rate of $|\cal X|$ can not be quadratic in $n$ or faster due to the converse result by  Kelly \cite{Kelly}.  This result states that,
in this regime, there is a probability that the supports of $T_1,T_2,...,T_M$, and the empirical distribution of $\vec{x}^n$ does not intersect and consistent classification is not possible. 

As an example, we have considered a large alphabet source, similar to the one in \cite[Thm. 4]{Kelly}, where $M=3$, $|{\cal X}| = n^{1.2}$ and the true distributions $P_i$, $i=1,2,3$, for the hypothesis are 
\begin{align*}
P_i(a_k) = \begin{cases} \frac{c}{|\cal X|} & \text{if} \quad  k=\frac{(i-1)|\cal X|}{M},\frac{(i-1)|\cal X|}{M}+1,...,\frac{i|\cal X|}{M},  \\
\frac{M-c}{(M-1) |\cal X|} & \text{otherwise},
\end{cases}
\end{align*}
where $1<c<M$ is a constant. For the  resultant distributions $\min_{i \neq j} V(P_i,P_j) = \frac{c-1}{M-1}$ for all $n$. In the simulation we have set $c=1.4$ to limit the running time.
The results are presented in Figure \ref{fig:fig_2} where we have also plotted the performance of the MAP test. For the considered sources, the DGL test started to perform consistently around $\alpha=1.6$ and the proposed upper bound becomes affective as $\sqrt{\alpha}$ gets comparable to $|\cal X|$. In this region, the performance of the proposed method gets close to that of MAP test, as well.

\section{Concluding Remarks} \label{sect:sect_5}

We have proposed to use the robust DGL test \cite{Devroye} for classification with labeled training sequences where the empirical distributions of the training sequences are used as nominal distributions. We have extended the non-asymptotic, exponential upper bound of the DGL test for the considered classification problem. 
The proposed upper bounds are simple to evaluate, but not tight in general. However, they can be useful for providing  lower bounds on the achievable error exponent in the non-asymptotic region. The proposed bounds can be further improved by tightening the error bound of the DGL test via Chernoff or Cramer-Rao Bound, as suggested in \cite{Biglieri}. When $M$ is not exponential in $n$, the proposed method has complexity $O(M^2n)$ which is quadratically larger, in the number of hypothesis, than optimal hypothesis testing. It can also be used if the alphabet size grows sub-quadratically in the length of the test sequence.

\vspace{-8 pt}
\section{APPENDIX}
\subsection{Proof of Propostion 1}
\vspace{-15 pt}
\vspace{-5 pt}
\begin{align}
 &\Pr[ \Phi^c]  = \Pr[  \cup_{i=1}^M \phi_i^c ]  = \sum_{i=1}^M 		\Pr[  \phi_i^c ] \nonumber \\ 		
                   &=\sum_{i}^M \Pr[ V(T_i,P_i) \geq \frac{\min_{i \neq j} V(T_i,T_j) -\Delta}{2}]  \label{eq:V_bound}\\
                  & = M \Pr[  \max_{F , F \subset {\cal X} } | T_i(F)-P_i(F) |  \geq  \frac{\min_{i \neq j} V(T_i,T_j) -\Delta}{2}] \nonumber  \\
                  &  \stackrel{a} {\leq}  M 2^{|{\cal X}|} \max_{F , F \subset {\cal X}} \Pr[ | T_i(F)-P_i(F) |  \geq  \frac{\min_{i \neq j} V(T_i,T_j) -\Delta}{2}] \nonumber \\
                   &  \stackrel{b} {\leq}   2Me^{-N \left( (\min_{i \neq j} V(T_i,T_j) -\Delta)^2/2-\frac{| {\cal X}| \ln 2}{N} \right)}\nonumber
\end{align}
where a) results from the union bound over $F, F \subset {\cal X}$; and b) results from Hoeffding's inequality \cite{Hoeffding}. Finally,
letting $N=n \alpha$ results in the desired bound.
\vspace{-10 pt}
\subsection{Proof of Propostion 2}
By double application of the triangle inequality we obtain
\begin{align*}
V(P_i,P_j) &\leq V( P_i,T_i) + V(T_i,P_j) \\
                 &\leq V( P_i,T_i) + V(P_j,T_j)+ V(T_i,T_j). 
\end{align*}
Since  $V(P_i,T_i) \leq \epsilon(\alpha)\min_{i \neq j} V(T_i,T_j) $ holds for all $i$ 
\begin{align*}
V(P_i,P_j) &\leq 2\epsilon(\alpha)\min_{i \neq j} V(T_i,T_j)  + V(T_i,T_j). 
\end{align*}
Taking $\min_{i  \neq j}$ of both sides results in the desired inequality.
\vspace{-10 pt}

\subsection{Proof of Propostion 3}
Using the definition \eqref{eq:V_def2} in \eqref{eq:V_bound} gives
\begin{align*}
 &\Pr[ \Phi^c]  = M \Pr[  \sum_{a \in \cal X} \frac{|T_i(a)-P_i(a)|}{2} \geq  \frac{\min_{i \neq j} V(T_i,T_j) -\Delta}{2}]  \\
                  &\leq   M \Pr[ | {\cal X} | \max_{a, a \in \cal X} |T_i(a)-P_i(a)| \geq  \min_{i \neq j} V(T_i,T_j) -\Delta  ]  \\
                  &\leq   M |{\cal X}|  \max_{a, a \in \cal X} | \Pr[ |T_i(a)-P_i(a)| \geq  \frac{\min_{i \neq j} V(T_i,T_j) -\Delta}{ | {\cal X} |}  ]  \\
                  &\leq   2M |{\cal X}|  e^{- N \left(  2 (\frac{\min_{i \neq j} V(T_i,T_j) -\Delta}{ | {\cal X} |})^2 \right) }, 
\end{align*}
where the last two inequalities result from the union bound over $a$, $a \in \cal X$, and Hoeffding's inequality, respectively. Letting $N=n \alpha$ concludes the proof.

\end{document}